\documentclass[aps,prl,twocolumn,groupedaddress,showpacs]{revtex4-1}
\usepackage{graphicx}
\usepackage{amssymb}
\usepackage{amsmath}
\usepackage{textcomp}
\usepackage{mathtools}
\usepackage{calrsfs}
\usepackage{psfrag}

\newcommand{\Ca}{{\rm Ca}}
\newcommand{\Caup}{{\rm Ca}_{\rm up}}

\newcommand{\keff}{\kappa_{\rm eff}}

\newcommand{\qbar}{\bar{q}}

\newcommand{\Hdot}{\dot{\mathcal{H}}}
\newcommand{\wdot}{\dot{w}}
\newcommand{\Wdot}{\dot{W}}
\newcommand{\wc}{\dot{w_c}}
\newcommand{\Wc}{\dot{W_c}}

\psfrag{Ca}[][][1.8]{$\Ca$}
\psfrag{Keff}[][][2.2]{$\keff$}
\psfrag{lambda}[][][2.2]{$\lambda$}
\psfrag{Timestep#}[][][1.8]{time step}
\psfrag{dtWminusD}[][][1.8]{$\Wdot, -D$}
\psfrag{Dc}[][][1.6]{$\Wc$}

\graphicspath{{figs/}}

\begin{document}

\title{Capillary Network Model:
Capillary Power and Effective Permeability}

\author{Morten Gr{\o}va}
\email{Morten.Grova@ntnu.no}

\affiliation{
Department of Physics,
Norwegian University of Science and Technology,
NO-7491 Trondheim,
Norway}

\date{May 5, 2012}

\begin{abstract}

A simple model of two-phase flow in porous media is presented. A
connection is made to statistical mechanics by applying capillary
power as a constraint. Stochastic sampling is then used to test the
validity of this approach. Good agreement is found between stochastic
sampling and time stepping for flow-rates above a transition value.

\end{abstract}

\pacs{47.56.+r, 47.61.Jd, 89.75.Da}

\maketitle

When one phase displaces another within a porous medium, complex
patterns are known to emerge \cite{lenormand1988numerical}. Less is
known about the flow patterns formed under steady-state conditions.
Dynamic effects, i.e., the dependence of the flow patterns on the
total flow-rate also remain poorly understood. This is despite the
significant importance to applications such as enhanced oil recovery,
groundwater contamination and water transport in fuel cells
\cite{dullien1992porous, sahimi1995flow, mukherjee2011pore}. Two-phase
flow in porous media also holds theoretical interest as a complex
system which exhibits self-organization. Self-organization is
evidenced by capillary pressure drops adding up to reduce effective
permeability rather than canceling out, even under steady-state
conditions.

Steady-state simulations have looked at the transport of disconnected
oil ganglia \cite{valavanides1998mechanistic} and relations between
driving pressure and fractional flow \cite{knudsen2002bulk}, amongst
other things. Experimental data can be classified by the model porous
medium: networks etched in glass \cite{avraam1994steady,
avraam1995generalized, avraam1995flow, avraam1999flow,
tsakiroglou2007transient}, Hele-Shaw cells \cite{tallakstad2009prl,
tallakstad2009pre} and bead packings \cite{rassi2011nuclear}.

Two recent experiments have explored the relationship between applied
pressure drop $\Delta P$ and flow-rate \cite{tallakstad2009prl,
tallakstad2009pre, rassi2011nuclear}. They have reported a power law
dependence, but have found different exponents. A two-dimensional
network simulator has been used to explore the proposed power law
\cite{grova2011two, grova2012two, sinha2012effective}. In the
simulator individual menisci are modeled; they are transported
according to the flow field, and created and destroyed in a manner
which crudely models snap-off and coalescence. Details are given in
\cite{grova2012two}.

\begin{figure}[t]
\begin{center}
  \scalebox{0.80}{\includegraphics{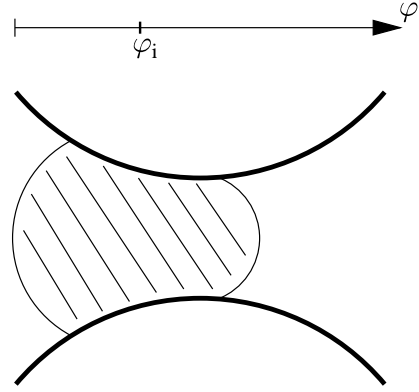}}
  \caption{
In the CNM, each link in the network is a capillary containing
a single droplet. The capillary models a pore throat; it is narrow
in the middle and wide near the pores. The position of the droplet
is given by $\varphi_i \in [0, 2\pi]$.
\label{fig:a}
}
\end{center}
\end{figure}

This paper presents a model which captures similar behavior as the
simulator without explicitly modeling menisci. The model consists of a
network of capillaries and will be referred to as the Capillary
Network Model (CNM). The capillaries model pore throats, which are the
narrow connections between pores. A single variable $\varphi$ is
assigned to each throat and a capillary pressure drop $p_c$ is given
as a function of $\varphi$.

The main control parameter used in simulations and experiments is the
capillary number $\Ca = \mu Q / \sigma \phi A$, where $\mu$ is
viscosity, $Q$ is volumetric flow-rate, $\sigma$ is surface tension,
$\phi$ is porosity and $A$ is the cross-sectional area of the sample.
The effective permeability is the sum of the relative permeabilities,
and may be defined as $\keff = Q / Q_0$ where $Q_0$ is the flow-rate
obtained by solving Darcy's law for single-phase flow with pressure
drop $\Delta P$. $\keff$ is found to be less than unity under
steady-state conditions, which signifies that the mixture of two
phases results in a larger resistance to flow than if only a single
phase is present.

In \cite{grova2011two, grova2012two} a modified version of the
Young-Laplace relation is used to obtain the capillary pressure drop
of a single meniscus
\begin{equation}
\label{eq:pc}
  p_c(x) = \frac{2\sigma}{r} \left[ 1-\cos(\frac{2\pi x}{\ell}) \right] ,
\end{equation}
where $x$ is the position of the meniscus, $\ell$ is the length of the
throat and $r$ measures its width. In the CNM the explicit menisci are
replaced by a single scalar variable for each pore throat. This
variable $\varphi$ is a coordinate which can take values between 0 and
$2\pi$. A capillary pressure drop function is defined as
\begin{equation}
\label{eq:cnm}
  p_c(\varphi) = \frac{4\sigma}{r} \sin(\pi s) \sin \varphi ,
\end{equation}
where $s$ is the non-wetting saturation of the throat, given by the
length of the droplet divided by the length of the throat. For
simplicity, all throats are assigned equal and constant $s$.
Eq.~\ref{eq:cnm} is obtained by considering Eq.~\ref{eq:pc} for the
case of two menisci forming a droplet within the pore throat, see
Fig.~\ref{fig:a}. $\varphi$ gives the position of the droplet. The
size of the droplet determines the non-wetting saturation $S$, and can
be removed as an independent parameter by redefining $\sigma$.

The porous medium is modeled as a 2D square lattice inclined at
$45^{\circ}$ relative to the main direction of flow. Boundary
conditions are bi-periodic, such that the network may be mapped onto a
torus. Flow is driven by a pressure drop applied across a cut through
the network.

Eq.~\ref{eq:cnm} models pore throats that have a narrowing geometry.
In all other respects pore throats are considered to be cylindrical
tubes with radius $r$. Network disorder may be introduced in $r$,
$\ell$ or both. In this work there is no disorder -- all pore throats
are equally wide and long. For all results given here, $r = 0.2 \ell$.

The Hagen-Poiseuille permeability for cylindrical tubes gives the
flow-rate $q$ by the Washburn equation,
\begin{equation}
\label{eq:washburn}
  q(\Delta p, \varphi) = -\frac{\pi r^4}{8\ell \mu} \left( \Delta p +
  p_c(\varphi) \right) ,
\end{equation}
where $\mu$ is the effective viscosity of the phases contained in the
throat and $\Delta p$ is the pressure difference between the two pores
connected by the throat.

Eqs.~\ref{eq:cnm} and \ref{eq:washburn} together with the network
geometry and Kirchhoff's circuit laws define the CNM.

\begin{figure}[t]
\begin{center}
  \scalebox{0.55}{\includegraphics{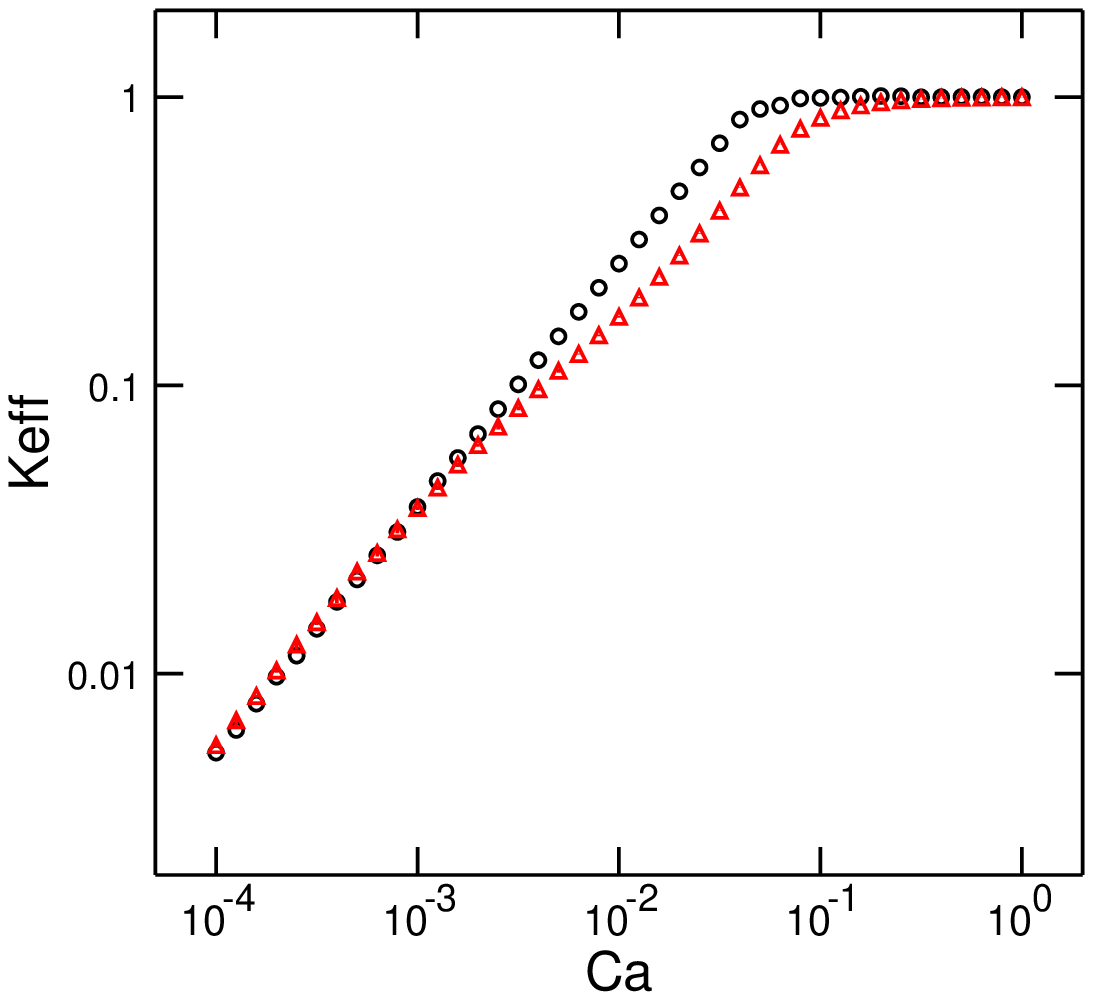}}
  \caption{
(Color online)
$\keff(\Ca)$ for the simulator (black circles) and the CNM (red
triangles). System size is 32x64.
\label{fig:c}
}
\end{center}
\end{figure}

Given a configuration consisting of $\{\varphi_i\}$ and $\Ca$, the
flow field is obtained by solving a system of linear equations. The
coordinates are then updated according to $\dot{\varphi_i} = q_i/a$
where $a = \pi r^2$. Numerical time stepping is done by the Euler
method, which is only first order accurate. The numerical error may be
estimated by comparing the capillary power with the total dissipation.
This ratio should be zero in the steady-state. It is found to be
negligible for $\Ca > 10^{-3}$, but grows to a few percent for $\Ca =
10^{-4}$. Below this the Euler method therefore does not provide
reliable results.

Fig.~\ref{fig:c} shows simulation results for the CNM and a network
simulator. In both cases Euler time stepping has been used. The network
simulator is the same as that used in \cite{grova2011two,
grova2012two}, but without disorder. Four independent runs underlie
each data point. Initialization is done by a random initial
configuration and a gradual increase in surface tension (gradually
decreasing $\Ca$). Saturation is a trivial parameter in the CNM, but
not in the simulator. The simulator results are for $S=0.4$. To
facilitate comparison between the CNM and the simulator, $s$ in
Eq.~\ref{eq:cnm} was set equal to $0.4$. The two models can be seen to
produce similar, but not identical results.

Consider Eq.~\ref{eq:washburn}. The right-hand side consists of two
terms, one proportional to $\Delta p$ and one to $p_c$. In obvious
notation $q = q_0 + q_c$. Multiplying with $q$ and dividing by $-\pi
r^4 / 8 \ell \mu$ gives a statement of conservation and conversion of
energy, $d = d_0 + d_c$, where $d$ is the heat dissipated through
viscous shear, always negative by definition, $-d_0 = -\Delta p~q$ is
equivalent to the power provided by a pump driving a flow $q$ with an
external pressure drop $\Delta p$ and $d_c = p_c q$ is capillary
dissipation. Replacing $-d_0$ with $\wdot$, where $w$ represents the
energy of the pump, and $-d_c$ with capillary power $\wc$ gives $\wdot
+ \wc + d = 0$. This states that the absolute value of heat
dissipation within a single throat is equal to the sum of applied
power and capillary power.

For the porous medium as a whole the capillary power adds up to $\Wc =
\sum \wc$. The pump power is $\Wdot = -\Delta P Q$ and the total heat
dissipation is $D = \sum d$. In general, the pump power and the heat
dissipation are not equal. The difference must be due to the
capillaries, so we have $\Wc = -\Wdot - D$ for the total capillary
power. If $\Wc$ is positive the capillary power adds to the pump power
to increase heat dissipation; it becomes an energy source. If $\Wc$ is
negative the capillary power absorbs some of the pump power to
decrease heat dissipation; it becomes an energy sink.

\begin{figure}[t]
\begin{center}
  \scalebox{0.60}{\includegraphics{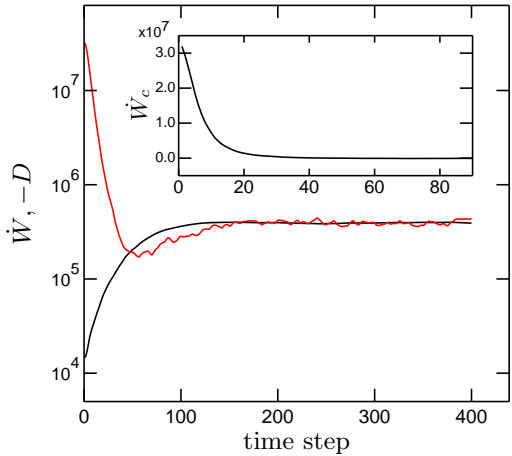}}
  \caption{
(Color online)
Time development of $\Wdot$ (black, initially increasing), $-D$ (red,
initially decreasing) and $\Wc$ (inset) for an initially random
configuration. The x-axis of the inset is a zoom-in compared to the
main figure. $\Ca = 10^{-3}$ is held constant and the system size is
128x256.
\label{fig:d}
}
\end{center}
\end{figure}

Fig.~\ref{fig:d} shows the development of $\Wdot$, $-D$ and $\Wc$
during a single simulation run. Initially, for a completely random
configuration, $\Wc$ is positive: the capillaries release energy.
$\Wc$ approaches zero as steady-state is approached, after which both
pump power and dissipation fluctuate around the same average value.

A thermodynamics of two-phase flow in porous media was first suggested
in \cite{hansen2009towards}. There, total dissipation was suggested as
being analogous to energy. The preceeding discussion implies that
instead of total dissipation it is capillary power which is
constrained and thus provides a connection to statistical mechanics.

In the following, tools from statistical mechanics usually reserved
for classical equilibrium will be applied. The underlying idea is to
consider the steady-state as governed by a balance between drive and
dissipation -- a dissipative equilibrium. $\langle \Wc \rangle = 0$
states that, on average, pump power must equal heat dissipation, which
is a requirement for the steady-state.

Constructing a space of eigenstates $\{\varphi_l\}$, where $l$ indexes
eigenstates, gives the number of microscopic configurations for a
state $\{a_l\}$ as
\begin{equation}
\label{eq:prob}
  C(\{a_l\}) = \frac{N!}{a_1!a_2!a_3! \dots} ,
\end{equation}
where $a_l$ is the occupancy number of eigenstate $l$ and $\sum a_l =
N$ is the number of coordinates \cite{schrodinger1952statistical}. In
light of the developments so far, and in order to establish the
governing principle of steady-state two-phase flow in porous media,
the capillary power $\Wc$ is applied as a constraint on
Eq.~\ref{eq:prob}. Using a Lagrange multiplier $\lambda$ for the
constraint gives the partition function
\begin{equation}
\label{eq:partfunc}
  Z = \sum_l e^{-\lambda \Hdot} ,
\end{equation}
where the sum runs over the eigenstates of $\varphi$. $\Hdot$ is
\begin{equation}
\label{eq:hamiltonian}
  \Hdot = -\sum_i p_c \qbar ,
\end{equation}
where the sum runs over the number of coordinates, $p_c =
p_c(\varphi_l)$ and $\qbar$ is
\begin{equation}
\label{eq:qbar}
  \qbar(\varphi_l) = \int dq ~ q ~ \rho(q|\varphi_l) .
\end{equation}

\begin{figure}[t]
\begin{center}
  \scalebox{0.58}{\includegraphics{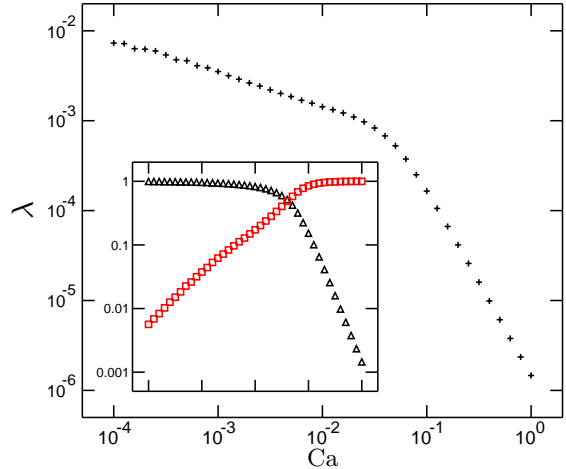}}
  \caption{
(Color online)
$\lambda(\Ca)$ for the CNM (plusses). Inset: $1-\keff$ (black
triangles) and $\keff$ (red squares) with the same x-axis as the main
figure. The data are from the same simulations as the CNM results from
Fig.~\ref{fig:c}.
\label{fig:x}
}
\end{center}
\end{figure}

The pdf's $\rho$ may be obtained from time stepping at a given $\Ca$.
Capillary power
\begin{equation}
\label{eq:Dc}
  \Wc(\lambda) = -Z^{-1} \partial_{\lambda} Z
  = -N Z^{-1} \sum_l p_c \qbar ~ e^{\lambda p_c \qbar} ,
\end{equation}
is a strictly monotonous function of $\lambda$, with only a single
solution $\lambda$ for a given value of $\Wc$. Fig.~\ref{fig:x} gives
$\lambda(\Ca)$ obtained by solving Eq.~\ref{eq:Dc} for $\Wc = 0$. A
transition occurs at $\Ca \approx 0.1$. This value is in the following
referred to as $\Caup$, to emphasize the connection with
\cite{grova2012two}.

The inset of Fig.~\ref{fig:x} shows that above $\Caup$, $(1-\keff)$
scales with $\Ca$. The best fit to a power law gives $(1-\keff) \sim
\Ca^{-2.06}$. A further discussion of scaling laws above $\Caup$ is
given in connection with the mean-field argument.

It should be noted that $\Wc(\lambda) = 0$ implies that
$\partial_{\lambda} Z = 0$, i.e., dissipative equilibrium implies that
the partition function has an extremal value with respect to
variations of the constraint. For $\Wc(\lambda) = 0$ to be valid, the
constraint must apply equally for each configuration within the
ensemble. Temporal correlations in the fluctuations of $\Wc$ are
ignored by this approach.

Using a Metropolis algorithm \cite{metropolis1953equation,
landau2009guide} and the value of $\lambda$ obtained from time stepping
it is possible to generate an ensemble of configurations governed by
the constraint of $\Wc=0$. The resulting ensemble will be dominated by
the most probable state (according to Eq.~\ref{eq:prob}) which
satisfies the constraint. A comparison with the ensemble obtained by
time stepping constitutes a non-trivial test of the validity of the
statistical mechanics approach.

Stochastic sampling is done by replacing some randomly chosen
coordinates with new, random coordinates. After a trial update,
$\Delta \Wc$ is calculated as the change in $\Wc$ after the update.
The new configuration is accepted if $\Delta \Wc$ is negative, and
with probability $e^{-\lambda \Delta \Wc}$ otherwise. This is a
standard Metropolis algorithm, where $\Wc$ is analogous to energy and
$\lambda$ is analogous to inverse temperature.

\begin{figure}[b]
\begin{center}
  \scalebox{0.60}{\includegraphics{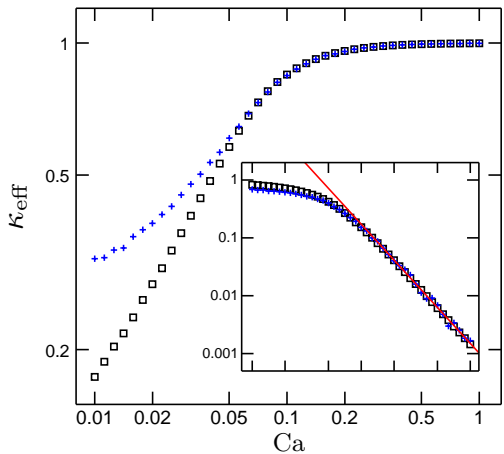}}
  \caption{
(Color online)
Comparison of $\keff (\Ca)$ obtained from time stepping (black squares)
and stochastic sampling (blue plusses). Inset: $1-\keff$ with the same
x-axis as the main figure. The solid red line is a power law with
exponent $-2.06$.
\label{fig:e}
}
\end{center}
\end{figure}

Fig.~\ref{fig:e} shows $\keff(\Ca)$ and $1-\keff(\Ca)$ from both
time stepping and stochastic sampling. For $\Ca$ above $\Caup$ the
results are in agreement. Below $\Caup$, time stepping and stochastic
sampling produce different results.

Inspection of time stepping simulations suggest that above $\Caup$,
$\qbar$ may be approximated by
\begin{equation}
\label{eq:mft}
  \qbar(\varphi) = q_0 - q_c \sin \varphi ,
\end{equation}
where $q_0$ is a mean flow-rate and $q_c$ is the maximum perturbation
caused by a capillary pressure drop. At $\Caup$, $q_c \approx q_0$.
Below $\Caup$ time stepping simulations show that the dependence of
$\qbar$ on $\varphi$ is no longer sinusoidal. Eq.~\ref{eq:mft}
represents a mean-field solution: a homogeneous flow field with
perturbations that only depend on the local variable.

Applying Eq.~\ref{eq:mft} to Eq.~\ref{eq:partfunc} allows an analytic
determination of the scaling of $\Ca$ and $(1-\keff)$ with $\lambda$.
To obtain this, the Boltzmann factor is expanded as $e^{-\lambda
\Hdot} \approx 1 - \lambda \Hdot$, omitting terms of $\lambda^2$ and
higher order. Terms with odd powers of $\sin \varphi$ sum to zero.
From $\Ca \sim \langle q \rangle$, $\Ca \sim q_0$ is obtained.
Considering $\langle \wc \rangle = 0$, $\Ca \sim \lambda^{-1/2}$
results. Finally, $\keff \sim \langle q \rangle^2/\langle q^2 \rangle$
gives $(1-\keff) \sim \lambda$. This gives $(1-\keff) \sim \Ca^{-2}$.

\begin{figure}[t]
\begin{center}
  \scalebox{0.55}{\includegraphics{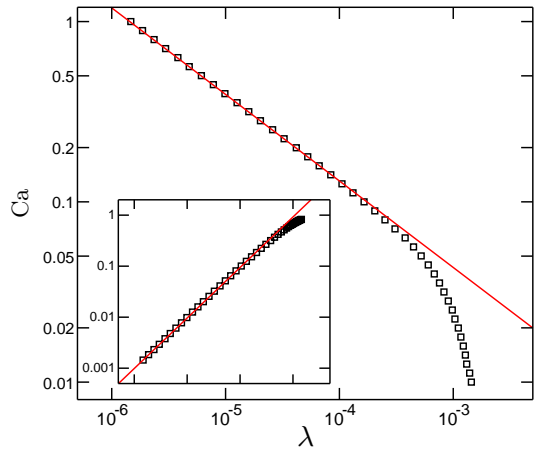}}
  \caption{
(Color online)
$\Ca(\lambda)$. The solid red line is a power law with exponent
$-0.48$. Inset: $1-\keff$ with the same x-axis as the main figure. The
solid red line is a power law with exponent $0.99$.
\label{fig:xx}
}
\end{center}
\end{figure}

From time stepping and above $\Caup$, the CNM gives $\Ca \sim
\lambda^{-0.48}$, $(1-\keff) \sim \lambda^{0.99}$ and $(1-\keff) \sim
\Ca^{-2.06}$, see the inset of Fig.~\ref{fig:e} and
Fig.~\ref{fig:xx}.

In summary, a simple model of two-phase flow in porous media has been
presented. It has been found to produce results for effective
permeability that are qualitatively similar to a more detailed
simulator. Capillary power has been defined and used as a constraint
to produce a partition function. At high flow-rates, time stepping and
stochastic sampling have been shown to produce the same ensemble,
given a constraint of zero capillary power. Scaling exponents obtained
from a mean-field theory are in agreement with the numerical results
at these high flow-rates. At lower flow-rates, stochastic sampling does
not reproduce the results from time stepping.

The main result of this work is to identify zero capillary power as
the constraint which governs steady-state two-phase flow in porous
media. This is nothing but a statement of conservation of energy.

Discussions with S.~Sinha and G.~T{\o}r{\aa} are gratefully
acknowledged.

\end{document}